\PassOptionsToPackage{unicode}{hyperref}
\PassOptionsToPackage{hyphens}{url}
\PassOptionsToPackage{dvipsnames,svgnames,x11names}{xcolor}
\documentclass[
  10pt,
]{article}
\usepackage{xcolor}
\usepackage[margin=0.82in]{geometry}
\usepackage{amsmath,amssymb}
\usepackage{iftex}
\ifPDFTeX
  \usepackage[T1]{fontenc}
  \usepackage[utf8]{inputenc}
  \usepackage{textcomp} 
\else 
  \usepackage{unicode-math} 
  \defaultfontfeatures{Scale=MatchLowercase}
  \defaultfontfeatures[\rmfamily]{Ligatures=TeX,Scale=1}
\fi
\usepackage{lmodern}
\ifPDFTeX\else
\fi
\IfFileExists{upquote.sty}{\usepackage{upquote}}{}
\IfFileExists{microtype.sty}{
  \usepackage[]{microtype}
  \UseMicrotypeSet[protrusion]{basicmath} 
}{}
\usepackage{setspace}
\makeatletter
\@ifundefined{KOMAClassName}{
  \IfFileExists{parskip.sty}{%
    \usepackage{parskip}
  }{
    \setlength{\parindent}{0pt}
    \setlength{\parskip}{6pt plus 2pt minus 1pt}}
}{
  \KOMAoptions{parskip=half}}
\makeatother
\usepackage{color}
\usepackage{fancyvrb}

\DefineVerbatimEnvironment{Highlighting}{Verbatim}{commandchars=\\\{\}}
\newenvironment{Shaded}{}{}

\newcommand{\AttributeTok}[1]{\textcolor[rgb]{0.49,0.56,0.16}{#1}}

\newcommand{\DataTypeTok}[1]{\textcolor[rgb]{0.56,0.13,0.00}{#1}}

\newcommand{\ExtensionTok}[1]{#1}

\newcommand{\FunctionTok}[1]{\textcolor[rgb]{0.02,0.16,0.49}{#1}}

\newcommand{\NormalTok}[1]{#1}

\usepackage{longtable,booktabs,array}
\usepackage{caption}
\usepackage{calc} 
\usepackage{etoolbox}
\makeatletter
\patchcmd\longtable{\par}{\if@noskipsec\mbox{}\fi\par}{}{}
\makeatother
\IfFileExists{footnotehyper.sty}{\usepackage{footnotehyper}}{\usepackage{footnote}}
\makesavenoteenv{longtable}
\usepackage{graphicx}
\makeatletter
\newsavebox\pandoc@box
\newcommand*\pandocbounded[1]{
  \sbox\pandoc@box{#1}%
  \Gscale@div\@tempa{\textheight}{\dimexpr\ht\pandoc@box+\dp\pandoc@box\relax}%
  \Gscale@div\@tempb{\linewidth}{\wd\pandoc@box}%
  \ifdim\@tempb\p@<\@tempa\p@\let\@tempa\@tempb\fi
  \ifdim\@tempa\p@<\p@\scalebox{\@tempa}{\usebox\pandoc@box}%
  \else\usebox{\pandoc@box}%
  \fi%
}
\def\fps@figure{htbp}
\makeatother
\providecommand{\tightlist}{%
  \setlength{\itemsep}{0pt}\setlength{\parskip}{0pt}}
\usepackage{bookmark}
\IfFileExists{xurl.sty}{\usepackage{xurl}}{} 
\makeatletter
\@ifundefined{xmpquote}{}{}
\makeatother
\hypersetup{
  pdftitle={RosettaBitcoin: An Artifact-Backed Experience Report on Verification Infrastructure for Agent-Assisted Consensus Validators},
  pdfauthor={Donavon Guyot --- Independent Researcher},
  colorlinks=true,
  linkcolor={RoyalBlue},
  filecolor={Maroon},
  citecolor={RoyalBlue},
  urlcolor={RoyalBlue},
  pdfcreator={LaTeX via pandoc}}

\title{RosettaBitcoin: An Artifact-Backed Experience Report on
Verification Infrastructure for Agent-Assisted Consensus Validators}
\author{Donavon Guyot --- Independent Researcher}
\date{August 2026}

\begin{document}
\maketitle

\setstretch{1.02}
\subsection{Abstract}\label{abstract}

Agent-assisted software projects are often reported through
demonstrations or aggregate benchmarks that conceal how correctness
claims were admitted. This experience report studies RosettaBitcoin, a
single-developer project that built twelve separately implemented
Bitcoin testnet4 consensus validators, through its immutable 17 June
2026 software snapshot (DOI
\href{https://doi.org/10.5281/zenodo.20738249}{\texttt{10.5281/zenodo.20738249}}).
We analyze the snapshot's tracked SQLite evidence database, curated
artifact index, conformance fixtures, validation scripts, blocker
records, and version history. At the snapshot, all twelve ports had
port-owned 45/45 script-corpus proofs and strict 5,000-block baselines.
Nine had canonical clean 50,000-block, 100,000-block, and post-100,000
validation lanes. Java had one 19.86-second near-tip maintenance
artifact. No port had an empty-state-to-tip proof, and no port satisfied
the project's binary full-node gate; Docker and live-node capability
gaps remained.

The artifact history also records a 3 h 17 min 57 s Zig
scaffold-to-50,000 span, but the observed intervals describe
non-equivalent tasks and cannot estimate effort, productivity, or
causality. A separate diagnostic supplement preserves evidence that a
pure-Mojo cryptographic backend validated fresh state to height 100,000
and resumed to 140,234, agreed on a 45-case shadow comparison, rejected
six crafted invalid classes, and was killed by three targeted mutations.
That evidence is noncanonical, noncomparable, and class-bounded. The
case suggests that explicit failure records, fixtures, port-owned
proofs, and validating imports can make agent-assisted systems more
auditable. Controlled ablations and external replications are needed to
test whether such infrastructure causally improves development outcomes.

\textbf{Keywords:} software engineering experience report;
agent-assisted development; consensus validation; differential testing;
research artifacts

\subsection{1. Introduction}\label{introduction}

Software produced with coding agents can be easy to demonstrate and
difficult to audit. A transcript may show how code was proposed, but it
does not establish which executable claims survived testing, whether
evidence was comparable, or which limitations were known when a result
was reported. These problems become acute in consensus software, where
an incorrect acceptance decision can remain silent until an
implementation disagrees with the network.

RosettaBitcoin is a monorepo containing consensus-validator
implementations in C++, C\#, Elixir, Go, Java, Mojo, OCaml, Python,
Rust, Swift, TypeScript, and Zig. The project uses Bitcoin testnet4
block bytes, a shared fixture corpus, explicit failure records,
port-owned proof artifacts, schema validators, and a tracked SQLite
database called Project. Its declared end condition is stronger than any
result reported here: from empty local state, a node must reach and
maintain the testnet4 tip while independently validating every stored
connected block.

This paper does not claim that RosettaBitcoin met that condition. It
treats the immutable 17 June 2026 Zenodo snapshot as one observational
case and asks:

\begin{enumerate}
\def\labelenumi{\arabic{enumi}.}
\tightlist
\item
  \textbf{RQ1:} How does RosettaBitcoin encode failures and admit
  evidence?
\item
  \textbf{RQ2:} What conformance and validation outcomes did the twelve
  implementations demonstrate by the snapshot?
\item
  \textbf{RQ3:} What longitudinal development pattern is visible in the
  artifact history, and what cannot be inferred about effort or
  causality?
\item
  \textbf{RQ4:} What bounded evidence supports the pure-Mojo diagnostic,
  and what remains unproved?
\end{enumerate}

The contributions are (i) a reconstruction of the project's verification
substrate, (ii) a snapshot-bounded inventory of its port-owned results,
(iii) a qualified longitudinal account grounded in versioned artifacts,
and (iv) a separate, hash-verified supplement for diagnostics that were
not canonical Project evidence. The accompanying claim--evidence matrix
makes each numerical claim traceable to a query, artifact, or manifest
entry.

\subsection{2. Related Work}\label{related-work}

\textbf{Testing without a complete oracle.} The oracle problem concerns
the difficulty of determining whether a program's output is correct
{[}1{]}. Differential testing compares independently produced outputs
for the same input and has long been used to expose compiler defects
{[}2{]}. RosettaBitcoin uses a related pattern, but agreement is not
treated as proof: shared misconceptions, common fixtures, and common
dependencies can make multiple implementations fail together. Crafted
negative cases and raw-chain replay add distinct checks.

\textbf{Multiple implementations and independence.} N-version
programming relies on diverse implementations, yet controlled evidence
shows that independently developed versions can exhibit correlated
failures {[}3{]}. Here, ``independent'' is therefore used narrowly. Each
port contains separately implemented consensus code that executes its
own validation decisions. The ports are not independent experimental
replications: they share one developer, one repository, one byte source,
one fixture corpus, blocker knowledge, evidence infrastructure, and some
storage and cryptographic dependencies.

\textbf{Case-study and repository methodology.} Software-engineering
case-study guidelines emphasize a defined case, explicit data
collection, traceable chains of evidence, and threats to validity
{[}4{]}. Mining-software-repository research similarly distinguishes the
repository objects, purpose, method, and evaluation used to draw
evolutionary conclusions {[}5{]}. We follow those principles by freezing
the case boundary at a DOI-backed snapshot and separating observations
from hypotheses. Reproducibility research warns that source availability
alone does not ensure repeatability {[}6{]}; consequently, this study
reports exact queries, validators, checksums, and evidence
classifications.

\textbf{Agent-assisted software engineering.} ReAct interleaves
language-model reasoning with actions in an external environment
{[}7{]}, while Reflexion studies language feedback retained across
trials {[}8{]}. SWE-bench evaluates language models on repository-level
issue resolution {[}9{]}. These works motivate attention to execution
and feedback, but they do not establish the causal effect of
RosettaBitcoin's substrate. This case contains no controlled agent
comparison, prompt archive, invocation count, token accounting, or
intervention-time log.

Bitcoin's consensus rules and historical chain divergences make
validation a useful high-consequence setting {[}10--13{]}. Bitcoin Core
is used here as a source of testnet4 bytes and post-snapshot provenance,
not as an imported validity verdict inside a port.

\subsection{3. Case Context and Study
Design}\label{case-context-and-study-design}

\subsubsection{3.1 Case boundary}\label{case-boundary}

The study object is the Zenodo record published on 17 June 2026, not the
subsequently evolving worktree. The downloaded
\texttt{rosetta-bitcoin-v1.0.0.zip} matches the MD5 published by Zenodo
and has directory boundary \texttt{4ade801}, resolving to Git commit
\texttt{4ade801b…}. The Project database and current-evidence index also
match the corresponding files at that commit. Complete archive, commit,
and SHA-256 values are retained in the claim--evidence matrix and
supplement manifest.

The archive contains the software, shared fixtures, Project database,
curated current-evidence index, validation scripts, and 91 tracked
conformance-result files. It does not contain the ignored local Mojo
diagnostic directory. Those files are therefore preserved in a separate
companion package and never retroactively imported into Project.

\subsubsection{3.2 Evidence classes}\label{evidence-classes}

We classify claims before interpreting them:

{\def\LTcaptype{none} 
\begin{longtable}[]{@{}
  >{\raggedright\arraybackslash}p{(\linewidth - 2\tabcolsep) * \real{0.5000}}
  >{\raggedright\arraybackslash}p{(\linewidth - 2\tabcolsep) * \real{0.5000}}@{}}
\toprule\noalign{}
\begin{minipage}[b]{\linewidth}\raggedright
Class
\end{minipage} & \begin{minipage}[b]{\linewidth}\raggedright
Meaning in this study
\end{minipage} \\
\midrule\noalign{}
\endhead
\bottomrule\noalign{}
\endlastfoot
Project-canonical & Imported and selected by the frozen Project
database/current-evidence index. \\
Diagnostic supplement & Hash-preserved evidence outside Project;
explicitly noncomparable. \\
Historical provenance & Commit or manifest evidence about sequence, not
effort. \\
Observation & A direct, reproducible description of the archived
materials. \\
Hypothesis & A proposition requiring a future controlled or external
study. \\
\end{longtable}
}

The claim--evidence matrix records the supporting query or file,
timestamp, limitations, and reproducibility status. Numerical sentences
were audited against that matrix. No ignored runtime database or log is
used as principal evidence.

\subsubsection{3.3 Analysis procedure}\label{analysis-procedure}

For RQ1, we inspected the schema, importers, validators, shared rule
ledger, fixture manifest, and reporting commands. For RQ2, we ran
read-only Project reports and strict preflights against the frozen
database and reconciled their rows with \texttt{current\_evidence.json}.
For RQ3, we used tagged commits and a tracked historical-provenance
manifest, retaining timestamp spans only where their endpoints were
explicit. For RQ4, we copied the original diagnostics without byte
changes, generated SHA-256 metadata, validated required fields, and
tested that none appeared in Project current evidence.

\subsubsection{3.4 Human and AI roles}\label{human-and-ai-roles}

One human developer coordinated the repository, selected tasks, judged
failures, and admitted evidence. Development was agent-assisted.
Repository records identify Codex/GPT-5 for the Mojo work; Claude
assisted with writing. The snapshot does not provide complete prompts,
invocation counts, token costs, human intervention time, or controlled
trials across models. We therefore make no claim about model dominance,
unaided model failure, or the fraction of work attributable to a model
or the developer.

\subsection{4. Verification Substrate}\label{verification-substrate}

\begin{figure}
\centering
\includegraphics[width=0.96\linewidth,height=\textheight,keepaspectratio,alt={The RosettaBitcoin evidence pipeline. Raw blocks and runtime failures are reduced to provenance-rich blocker facts; reusable fixtures and rule cards are executed by each port; port-owned artifacts pass schema and gate validators before import; Project reports expose admitted evidence. Agreement is one check, not a substitute for negative tests or independent implementation.}]{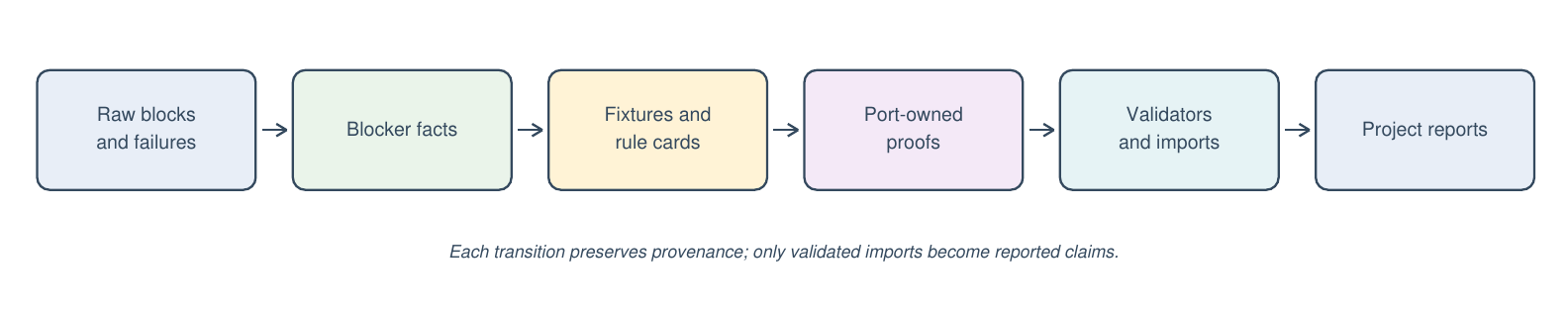}
\caption{The RosettaBitcoin evidence pipeline. Raw blocks and runtime
failures are reduced to provenance-rich blocker facts; reusable fixtures
and rule cards are executed by each port; port-owned artifacts pass
schema and gate validators before import; Project reports expose
admitted evidence. Agreement is one check, not a substitute for negative
tests or independent implementation.}
\end{figure}

The substrate converts runtime discoveries into bounded, reusable tests.
A blocker record is intended to contain height, block hash, transaction
ID, input index, spent output script, observed failure, missing rule,
implementing fix, fixture, and follower notes. The shared script
manifest contains 45 fixtures, and the rule inventory contains 45
corresponding rule records. The Project database contains 111 normalized
blocker rows. These counts describe archived objects; they do not show
that the corpus is complete.

Each port produces its own corpus and benchmark evidence. A shared
validator checks schema, storage/cryptography declarations, target
height, fresh-state conditions, telemetry, and other gate-specific
fields. Import scripts populate Project, and reports query Project
rather than reconstructing status from prose. This separation
distinguishes three propositions that are otherwise easy to conflate: an
implementation ran; its artifact satisfied a declared schema; and the
artifact was selected as current comparable evidence.

\subsection{5. Results}\label{results}

\subsubsection{5.1 RQ1: Encoding failures and admitting
evidence}\label{rq1-encoding-failures-and-admitting-evidence}

The archived process is a one-way evidentiary pipeline rather than an
informal collection of status notes. Raw-chain execution supplies a
concrete failure; the failure is normalized as blocker facts; those
facts can become a portable fixture and rule card; each port emits a
port-owned result; validators reject malformed or out-of-policy
artifacts; importers admit accepted rows; reports render the resulting
state. \texttt{current\_evidence.json} selects 63 current entries: 12
corpus, 12 baseline-5k, nine shakedown-50k, nine performance-100k, nine
post-100k, ten storage, one external-probe, and one tip-maintenance
entry.

This design provides traceability and makes some status contradictions
detectable. It does not guarantee semantic completeness. A wrong shared
rule, an inadequate negative corpus, a validator bug, or a developer's
mistaken admission decision could be propagated across ports.

\subsubsection{5.2 RQ2: Snapshot outcomes}\label{rq2-snapshot-outcomes}

All twelve ports passed the port-owned 45/45 corpus gate and strict 5k
baseline preflight. Nine active contenders also had canonical, clean
50k, 100k, and post-100k lanes. The remaining Python, Elixir, and
TypeScript ports were baseline-retired. Table 1 reports the frozen
Project posture.

{\def\LTcaptype{none} 
\begin{longtable}[]{@{}
  >{\raggedright\arraybackslash}p{(\linewidth - 6\tabcolsep) * \real{0.2143}}
  >{\raggedleft\arraybackslash}p{(\linewidth - 6\tabcolsep) * \real{0.2857}}
  >{\raggedleft\arraybackslash}p{(\linewidth - 6\tabcolsep) * \real{0.2857}}
  >{\raggedright\arraybackslash}p{(\linewidth - 6\tabcolsep) * \real{0.2143}}@{}}
\toprule\noalign{}
\begin{minipage}[b]{\linewidth}\raggedright
Port
\end{minipage} & \begin{minipage}[b]{\linewidth}\raggedleft
Snapshot lifecycle
\end{minipage} & \begin{minipage}[b]{\linewidth}\raggedleft
Highest Project-reported validated height
\end{minipage} & \begin{minipage}[b]{\linewidth}\raggedright
Highest canonical lane
\end{minipage} \\
\midrule\noalign{}
\endhead
\bottomrule\noalign{}
\endlastfoot
C++ & active contender & 138,504 & post-100k \\
C\# & active contender & 138,511 & post-100k \\
Elixir & baseline retired & 5,000 & baseline-5k \\
Go & active contender & 138,505 & post-100k \\
Java & active contender & 138,591 & post-100k + short maintenance \\
Mojo & active contender & 140,134 & post-100k \\
OCaml & active contender & 138,512 & post-100k \\
Python & baseline retired & 52,996 & baseline-5k \\
Rust & active contender & 138,505 & post-100k \\
Swift & active contender & 138,504 & post-100k \\
TypeScript & baseline retired & 5,000 & baseline-5k \\
Zig & active contender & 138,504 & post-100k \\
\end{longtable}
}

\emph{Table 1. Snapshot posture from Project reports. Every row also has
corpus 45/45 and a strict baseline-5k pass. ``Post-100k'' begins from
admitted 100k state and is not an empty-state-to-tip run.}

Java's selected maintenance artifact covers 19.86 seconds, from height
138,575 to 138,591. It is evidence of a short exercised interval, not
durable tip maintenance. There are zero \texttt{tip\_once} artifacts. No
port demonstrated the binary gate. Project also reported only partial
Docker/supervisor coverage and unclosed full-node capabilities,
including public-peer operation, serving, relay/mempool behavior,
reorganization handling, crash recovery, restart soak, and
adversarial/resource-safety work. Thus the snapshot supports a set of
consensus-validation milestones, not twelve completed Bitcoin nodes.

The separate codebases make their own validation decisions, so
``independent'' is useful at the implementation boundary. It must not be
read as experimental independence: the common developer, knowledge base,
fixtures, byte source, and infrastructure make correlated error
plausible.

\subsubsection{5.3 RQ3: Longitudinal pattern and its
limits}\label{rq3-longitudinal-pattern-and-its-limits}

The historical record contains successive artifact eras:

{\def\LTcaptype{none} 
\begin{longtable}[]{@{}
  >{\raggedright\arraybackslash}p{(\linewidth - 4\tabcolsep) * \real{0.3333}}
  >{\raggedright\arraybackslash}p{(\linewidth - 4\tabcolsep) * \real{0.3333}}
  >{\raggedright\arraybackslash}p{(\linewidth - 4\tabcolsep) * \real{0.3333}}@{}}
\toprule\noalign{}
\begin{minipage}[b]{\linewidth}\raggedright
Recorded span
\end{minipage} & \begin{minipage}[b]{\linewidth}\raggedright
Endpoint artifacts
\end{minipage} & \begin{minipage}[b]{\linewidth}\raggedright
Interpretation permitted here
\end{minipage} \\
\midrule\noalign{}
\endhead
\bottomrule\noalign{}
\endlastfoot
20 Apr--4 May 2026 & 30,549-line assertion intermediate language; no
running node & A 14-calendar-day formalization span ended without the
later product artifact. \\
3--23 May 2026 & coordination/provenance infrastructure; no running node
& A 20-calendar-day process span preceded the product pivot. \\
23 May--8 Jun 2026 & Java product-track commits & A 16-calendar-day span
contains the lead implementation's recorded development. \\
5 Jun 2026, 12:38--15:56 & Zig empty scaffold to 50k shakedown & Commit
timestamps delimit 3 h 17 min 57 s. \\
\end{longtable}
}

\emph{Table 2. Non-equivalent artifact and commit spans. The rows differ
in goals, starting assets, execution time, review, and accumulated
shared knowledge.}

The visible pattern is that later ports appear after more shared
fixtures, blocker knowledge, and automation existed. The Zig interval is
a reproducible repository observation, but it is not a controlled effort
measure. Build and sync work can occur between commits; inactive time is
included; earlier and later tasks are not equivalent; agent invocations
and human interventions were not recorded. The history therefore
supports neither an inferred cost ratio nor a causal productivity
effect. It motivates a future hypothesis: under a fixed task and agent
configuration, a pre-populated verification substrate may reduce time to
admitted evidence relative to an otherwise identical empty substrate.

\subsubsection{5.4 RQ4: Pure-Mojo
diagnostic}\label{rq4-pure-mojo-diagnostic}

The companion archive (DOI
\href{https://doi.org/10.5281/zenodo.22114337}{\texttt{10.5281/zenodo.22114337}})
uses schema \texttt{rosettabitcoin.paper\_supplement.v1}. Every entry
records the original path, SHA-256, backend, supported claim, and
\texttt{does\_not\_prove} boundaries; the package is globally marked
\texttt{diagnostic\_non\_comparable} and
\texttt{project\_current\_evidence=false}.

{\def\LTcaptype{none} 
\begin{longtable}[]{@{}
  >{\raggedright\arraybackslash}p{(\linewidth - 4\tabcolsep) * \real{0.3333}}
  >{\raggedright\arraybackslash}p{(\linewidth - 4\tabcolsep) * \real{0.3333}}
  >{\raggedright\arraybackslash}p{(\linewidth - 4\tabcolsep) * \real{0.3333}}@{}}
\toprule\noalign{}
\begin{minipage}[b]{\linewidth}\raggedright
Diagnostic
\end{minipage} & \begin{minipage}[b]{\linewidth}\raggedright
Recorded result
\end{minipage} & \begin{minipage}[b]{\linewidth}\raggedright
Boundary
\end{minipage} \\
\midrule\noalign{}
\endhead
\bottomrule\noalign{}
\endlastfoot
Pure 5k & validation to 5,000 & diagnostic, not a baseline comparison \\
Pure fresh-state 100k & 0 to 100,000 & not the workspace binary gate \\
Pure resume & 100,000 to 140,234 & not empty-state-to-tip \\
Shadow corpus & 45 recorded agreements & limited to the shared fixture
set \\
Pure/native reject checks & six invalid families rejected by each
backend & class-bounded negative testing \\
Fault injection & ECDSA-, Schnorr-, and Taproot-tweak-accepting
mutations each made the check fail & three mutation classes, not
mutation adequacy in general \\
\end{longtable}
}

\emph{Table 3. Bounded diagnostic evidence preserved outside Project.}

The validation artifacts identify \texttt{mojo-pure-secp256k1}, declare
no native cryptographic backend, and report
\texttt{native\_fallback\_used=false}. The six reject families cover
P2PKH/ECDSA, bare multisig, P2SH binding, SegWit-v0 witness binding,
Taproot tweak, and Taproot Schnorr cases. The three fault modes caused
two, one, and one invalid cases, respectively, to be accepted, so the
test turned red in each targeted mode.

One consensus blocker can now be reported with full provenance, but only
as post-snapshot supplemental evidence. At height 56,447, block
\texttt{000000000ae11eebf9807d5ac0f9966282c59a1e613e229aa2bf7aea266d4535},
transaction
\texttt{cb5fe6b28e78371a6fd7f9439542ff6bf0082f070097c9a096220f2a786d521f},
input 0 spends a legacy bare 1-of-3 multisig output containing malformed
candidate public keys. The recovered rule is that a malformed public key
in legacy non-witness signature matching is a failed match rather than a
fatal script error; stricter witness-v0 and DER behavior remains
separate. This recovery improves traceability but does not move the
snapshot's gates.

These observations do not establish production cryptographic safety,
side-channel resistance, full Bitcoin consensus coverage, benchmark
comparability, model capability, or full-node completion. Pure Mojo is a
secondary diagnostic, not the study's principal result.

\subsection{6. Discussion}\label{discussion}

The case's most transferable artifact is the evidence architecture, not
a port count or a speed claim. The pipeline gives failures a stable
identity, converts selected failures into executable tests, requires
every implementation to own its result, and separates artifact validity
from evidence selection. For an agent-assisted project, those properties
create external, queryable project memory without asking a model or
developer to reconstruct status from prose.

The multiple ports add useful heterogeneity, but shared infrastructure
creates common-mode risk. Agreement across languages can increase
confidence only when the decision paths are genuinely separate and the
tests can expose false acceptance as well as false rejection. The
negative corpus and mutation checks are steps toward that goal, not
evidence that the goal is complete.

For practitioners, the case suggests three design questions: Can a
failure be recorded with enough provenance to replay? Does every claimed
implementation emit its own machine-checkable proof? Is current status
generated from admitted evidence rather than maintained in narrative
text? These are engineering questions. Whether answering them improves
delivery time or defect discovery is an empirical question this study
cannot resolve.

\subsubsection{Future work}\label{future-work}

The outstanding node roadmap remains future work: empty-state-to-tip
proof, meaningful tip maintenance, complete Docker/supervisor
conformance, serving and relay behavior, reorganization and recovery
tests, and adversarial/resource safety. The shared negative corpus
should expand beyond its current classes. The verification-substrate
hypothesis requires controlled ablation with fixed tasks, environments,
human policies, and agent configurations. External teams should
replicate both the evidence pipeline and selected validators without
sharing the original developer's tacit knowledge. These activities
extend the case; they are not prerequisites for interpreting the frozen
artifact report.

\subsection{7. Threats to Validity}\label{threats-to-validity}

\textbf{Construct validity.} Heights, fixture counts, and admitted
artifacts measure declared gates, not complete node correctness. A
45-case corpus cannot cover all Bitcoin scripts. ``Validated height''
depends on the implementation and evidence schema behaving as intended.
The study avoids treating port count as a direct measure of reliability.

\textbf{Internal validity.} This is not an experiment. Infrastructure,
developer experience, agent behavior, available hardware, repository
contents, and task selection all changed over time. Commit intervals are
not effort logs. No causal productivity or model-capability conclusion
follows.

\textbf{External validity.} One developer, one repository, one protocol,
and one test network limit generalization. Consensus validation has
unusually concrete inputs and agreement conditions. Other domains may
have slower or less observable feedback.

\textbf{Reliability and common-mode error.} The DOI and hashes stabilize
the study object, and the matrix makes the analysis auditable.
Nevertheless, the author also built the system, selected its evidence,
and wrote this report. Ports share fixtures, blocker interpretations,
bytes, and dependencies. There was no independent reproduction. The
separate supplement was assembled after the snapshot; byte hashes
protect its copied files but cannot retroactively make them part of the
original record.

\subsection{8. Artifact Availability and
Reproduction}\label{artifact-availability-and-reproduction}

The authoritative software snapshot is
\href{https://doi.org/10.5281/zenodo.20738249}{\texttt{10.5281/zenodo.20738249}}.
The repository contains the Markdown source, claim--evidence matrix,
figure source, build script, and companion diagnostic package source.
The diagnostic supplement is archived separately at DOI
\href{https://doi.org/10.5281/zenodo.22114337}{\texttt{10.5281/zenodo.22114337}}
and is related to---not incorporated into---the original snapshot.

Read-only snapshot checks:

\begin{Shaded}
\begin{Highlighting}[]
\ExtensionTok{python3}\NormalTok{ Project/scripts/report.py }\AttributeTok{{-}{-}db}\NormalTok{ Project/project.db }\AttributeTok{{-}{-}section}\NormalTok{ port{-}status}
\ExtensionTok{python3}\NormalTok{ Project/scripts/report.py }\AttributeTok{{-}{-}db}\NormalTok{ Project/project.db }\AttributeTok{{-}{-}section}\NormalTok{ benchmark{-}suite}
\ExtensionTok{python3}\NormalTok{ Project/scripts/report.py }\AttributeTok{{-}{-}db}\NormalTok{ Project/project.db }\AttributeTok{{-}{-}section}\NormalTok{ consensus{-}runway}
\ExtensionTok{python3}\NormalTok{ Project/scripts/preflight\_port\_baseline.py }\DataTypeTok{\textbackslash{}}
  \AttributeTok{{-}{-}db}\NormalTok{ Project/project.db }\AttributeTok{{-}{-}all} \AttributeTok{{-}{-}strict}
\ExtensionTok{python3}\NormalTok{ Project/scripts/preflight\_consensus\_runway.py }\DataTypeTok{\textbackslash{}}
  \AttributeTok{{-}{-}db}\NormalTok{ Project/project.db }\AttributeTok{{-}{-}all} \AttributeTok{{-}{-}stage}\NormalTok{ corpus }\AttributeTok{{-}{-}strict}
\end{Highlighting}
\end{Shaded}

Supplement verification and deterministic packaging:

\begin{Shaded}
\begin{Highlighting}[]
\ExtensionTok{python3}\NormalTok{ Docs/paper/supplement/verify.py}
\ExtensionTok{python3}\NormalTok{ Docs/paper/supplement/build.py}
\end{Highlighting}
\end{Shaded}

These commands verify archived JSON and package metadata; they do not
regenerate the diagnostic runs. Build the paper and arXiv bundle with:

\begin{Shaded}
\begin{Highlighting}[]
\FunctionTok{bash}\NormalTok{ Docs/paper/build.sh}
\end{Highlighting}
\end{Shaded}

\subsection{9. Conclusion}\label{conclusion}

At its immutable June 17 boundary, RosettaBitcoin had twelve port-owned
corpus and 5k proofs, nine clean canonical long-run lanes, one very
short maintenance artifact, and no empty-state-to-tip result. A
separate, noncanonical supplement supports a bounded pure-Mojo
diagnostic through height 140,234 and targeted negative checks. The case
shows how an agent-assisted project can encode failures, validate
evidence, and expose limitations in reproducible artifacts. It does not
show that the infrastructure caused faster development or that any model
can autonomously build a complete node. A controlled substrate ablation
and an external replication are the appropriate tests of that future
claim.

\subsection{References}\label{references}

{[}1{]} E. T. Barr, M. Harman, P. McMinn, M. Shahbaz, and S. Yoo. ``The
Oracle Problem in Software Testing: A Survey.'' \emph{IEEE Transactions
on Software Engineering} 41(5), 507--525, 2015.
\href{https://doi.org/10.1109/TSE.2014.2372785}{\texttt{10.1109/TSE.2014.2372785}}.

{[}2{]} W. M. McKeeman. ``Differential Testing for Software.''
\emph{Digital Technical Journal} 10(1), 100--107, 1998.
\href{https://vmssoftware.com/docs/dtj-v10-01-1998.pdf}{Primary journal
PDF}.

{[}3{]} J. C. Knight and N. G. Leveson. ``An Experimental Evaluation of
the Assumption of Independence in Multiversion Programming.'' \emph{IEEE
Transactions on Software Engineering} SE-12(1), 96--109, 1986.
\href{https://doi.org/10.1109/TSE.1986.6312924}{\texttt{10.1109/TSE.1986.6312924}}.

{[}4{]} P. Runeson and M. Höst. ``Guidelines for Conducting and
Reporting Case Study Research in Software Engineering.'' \emph{Empirical
Software Engineering} 14, 131--164, 2009.
\href{https://doi.org/10.1007/s10664-008-9102-8}{\texttt{10.1007/s10664-008-9102-8}}.

{[}5{]} H. Kagdi, M. L. Collard, and J. I. Maletic. ``A Survey and
Taxonomy of Approaches for Mining Software Repositories in the Context
of Software Evolution.'' \emph{Journal of Software Maintenance and
Evolution} 19(2), 77--131, 2007.
\href{https://doi.org/10.1002/smr.344}{\texttt{10.1002/smr.344}}.

{[}6{]} C. Collberg and T. A. Proebsting. ``Repeatability in Computer
Systems Research.'' \emph{Communications of the ACM} 59(3), 62--69,
2016. \href{https://doi.org/10.1145/2812803}{\texttt{10.1145/2812803}}.

{[}7{]} S. Yao et al.~``ReAct: Synergizing Reasoning and Acting in
Language Models.'' \emph{ICLR}, 2023.
\href{https://arxiv.org/abs/2210.03629}{\texttt{arXiv:2210.03629}}.

{[}8{]} N. Shinn et al.~``Reflexion: Language Agents with Verbal
Reinforcement Learning.'' \emph{NeurIPS}, 2023.
\href{https://arxiv.org/abs/2303.11366}{\texttt{arXiv:2303.11366}}.

{[}9{]} C. E. Jimenez et al.~``SWE-bench: Can Language Models Resolve
Real-World GitHub Issues?'' \emph{ICLR}, 2024.
\href{https://arxiv.org/abs/2310.06770}{\texttt{arXiv:2310.06770}}.

{[}10{]} S. Nakamoto. ``Bitcoin: A Peer-to-Peer Electronic Cash
System.'' 2008.
\href{https://bitcoin.org/bitcoin.pdf}{\texttt{bitcoin.org/bitcoin.pdf}}.

{[}11{]} Bitcoin Core developers. ``BIP 50: March 2013 Chain Fork
Post-Mortem.'' 2013. \href{https://bips.dev/50/}{\texttt{bips.dev/50}}.

{[}12{]} P. Wuille. ``BIP 66: Strict DER Signatures.'' 2015.
\href{https://bips.dev/66/}{\texttt{bips.dev/66}}.

{[}13{]} F. Jahr. ``BIP 94: Testnet 4.'' 2024.
\href{https://bips.dev/94/}{\texttt{bips.dev/94}}.

{[}14{]} D. Guyot. \emph{RosettaBitcoin v1.0.0}. Zenodo, 17 June 2026.
\href{https://doi.org/10.5281/zenodo.20738249}{\texttt{10.5281/zenodo.20738249}}.

{[}15{]} D. Guyot. \emph{RosettaBitcoin Mojo Diagnostic Supplement:
Pure-Backend Validation and Negative Tests}. Zenodo, version 1, 2026.
\href{https://doi.org/10.5281/zenodo.22114337}{\texttt{10.5281/zenodo.22114337}}.

\end{document}